\documentclass[]{spie}  

\usepackage{amsmath,amsfonts,amssymb}
\usepackage{graphicx}
\usepackage[colorlinks=true, allcolors=blue]{hyperref}

\title{The Missing MeV Window for Identifying Galactic PeVatrons: the need for a dedicated MeV Observatory}

\author[a]{Martina Cardillo}
\author[b]{Riccardo Campana}
\author[a]{Yuri Evangelista}
\author[a]{Fabio Muleri}
\affil[a]{INAF-IAPS, Via del Fosso del Cavaliere 100, 00133, Roma, Italia}
\affil[b]{INAF-OAS, via Gobetti 93/3 40129, Bologna, Italia}

\authorinfo{Further author information:\\Martina Cardillo: E-mail: martina.cardillo@inaf.it, Telephone: 0039 0649934462}

\begin{document} 
\maketitle

\begin{abstract}

The identification of the sources responsible for accelerating Galactic cosmic-rays up to PeV energies remains one of the major open questions in high-energy astrophysics. Recent observations by the Large High Altitude Air Shower Observatory (LHAASO) have revealed a rapidly growing population of Galactic PeVatron candidates opening a new era in the study of extreme particle accelerators. However, the detection of TeV-PeV gamma-rays alone does not uniquely identify the nature of the emitting particles. In many cases, both hadronic and leptonic scenarios can reproduce the observed very-high-energy emission, leaving the origin of the radiation unresolved.

The most robust and least model-dependent electromagnetic tracer of hadronic acceleration is provided by neutral-pion decay, which produces the characteristic \textit{pion bump} in the 10-100 MeV energy range. Observations in this energy band are therefore essential to establish whether candidate PeVatrons are accelerating protons and nuclei and, consequently, to assess their role in the origin of Galactic cosmic-rays. More broadly, the MeV domain addresses a wide range of key astrophysical questions, but despite its scientific importance, this energy range remains one of the least explored regions of the electromagnetic spectrum. As current and future VHE facilities will continue to expand the Galactic PeVatron census, the lack of sensitive MeV observations is becoming a major limitation for source identification.

We discuss the astrophysical science drivers and observational requirements for a dedicated MeV gamma-ray capability optimized for \textit{pion-bump} studies. We explore the concept of a compact, focused, and potentially fast-track mission capable of filling the impending MeV observational gap while more ambitious next-generation MeV facilities are being developed and evaluated. Such a mission would provide the missing diagnostic needed to distinguish hadronic and leptonic emission, identify Galactic cosmic-ray accelerators, and fully exploit the scientific return of the emerging generation of TeV observatories.

\end{abstract}


\section{INTRODUCTION}
\label{sec:intro}  

The identification of the sources responsible for accelerating Galactic cosmic-rays (CRs) up to PeV energies remains one of the longest-standing open problems in high-energy (HE) astrophysics. For decades, supernova remnants (SNRs) have been considered the most promising candidates, yet direct observational evidence capable of unambiguously demonstrating hadronic acceleration up to the CR knee (E$\sim3\times10^{15}$ eV) has remained elusive. In recent years, the situation has evolved rapidly thanks to the advent of a new generation of very-high-energy (VHE) gamma-ray observatories. In particular, the Large High Altitude Air Shower Observatory (LHAASO) has revealed a growing population of Galactic sources emitting photons up to hundreds of TeV and, in some cases, approaching the PeV domain, establishing new classes of candidate Galactic PeVatrons, i.e. sources capable of accelerating particles up to PeV energies \cite{Cao21,Cao24_LHAASO24_cat}: pulsar wind nebulae (PWNe), young massive stellar clusters (YMCs) and microquasars (MQs). Future facilities such as the Cherenkov Telescope Array Observatory (CTAO) \cite{CTA19} and the ASTRI Mini-Array \cite{Scuderi22,Vercellone22} are expected to dramatically expand this population and provide unprecedented views of the non-thermal Galactic sky.

These discoveries mark a fundamental transition in the study of extreme particle accelerators. For the first time, Galactic PeVatrons are becoming a population science rather than the investigation of a handful of exceptional objects. Since Galactic CRs are predominantly hadronic, establishing the hadronic nature of these sources is a prerequisite for determining their contribution to the Galactic CR population and, ultimately, for resolving the origin of Galactic CRs. However, detecting a PeVatron candidate is not equivalent to identifying a CR accelerator. While TeV-PeV gamma-rays reveal the presence of extreme particle acceleration, they do not uniquely determine the nature of the accelerated particles. In many cases, both hadronic and leptonic scenarios can reproduce the observed VHE emission, particularly in complex environments where pulsars, PWNe, SNRs, molecular clouds (MCs), and YMCs coexist. As a result, the nature of several candidate PeVatrons remains uncertain despite the remarkable progress achieved at the highest energies.

A definitive confirmation of hadronic acceleration would ideally be provided by the coincident detection of gamma-rays and HE neutrinos from the same source. Neutrinos provide the only direct probe of charged-pion production and therefore represent the most unambiguous signature of hadronic interactions. While recent multi-messenger observations have demonstrated the power of this approach for extragalactic accelerators, the expected neutrino fluxes from most Galactic PeVatron candidates remain below the sensitivity of current facilities, making source-by-source identification challenging in the foreseeable future \cite{AdrianMartinez16-KM3}.

In this context, neutral-pion decay provides the most direct electromagnetic tracer of hadronic particle acceleration. The resulting spectral feature, commonly known as the \textit{pion bump}, falls in the poorly explored 10-100 MeV energy range and offers a unique diagnostic for identifying candidate CR accelerators \cite{Giuliani11,Ackermann13,Liu24_MeV}. Unlike broadband TeV-PeV emission, whose interpretation often relies on detailed modeling of particle populations and source environments, the \textit{pion bump} is directly linked to hadronic interactions and therefore provides a much more robust signature of relativistic protons and nuclei.

The importance of the MeV domain extends well beyond the study of Galactic PeVatrons. This energy range plays a central role in a broad range of astrophysical investigations, including CR acceleration and transport, diffuse Galactic gamma-ray emission, CR interactions with the interstellar medium, compact objects and transients, and multi-messenger astrophysics. Despite its scientific importance, the MeV domain remains one of the least explored regions of the electromagnetic spectrum. Although several mission concepts targeting this energy range have been proposed in recent years, including e-ASTROGAM \cite{DeAngelis18,DeAngelis_26}, AMEGO-X \cite{McEnery19_AMEGO}, ComPair \cite{Moissev15_ComPair,Valverde23_ComPair} and the only accepted one COSI \cite{Tomsick19_COSI}, no approved mission currently provides the combination of sensitivity and energy coverage required to systematically investigate \textit{pion-bump} signatures in Galactic PeVatron candidates. As the census of candidate PeVatrons continues to grow, this lack of observational capability is becoming a major limitation for source identification and for our understanding of the origin of Galactic CRs.

In this work, we discuss the astrophysical science drivers for a dedicated MeV gamma-ray capability optimized for \textit{pion-bump} studies. We explore the concept of a compact and focused mission capable of filling the impending MeV observational gap while more ambitious next-generation MeV facilities are being developed and evaluated. Before discussing the observational requirements for such measurements, it is important to understand why the identification of hadronic PeVatrons has become a central issue in the broader quest to determine the origin of Galactic CRs.

\begin{figure}[ht]
\centering
\includegraphics[width=0.6\textwidth]{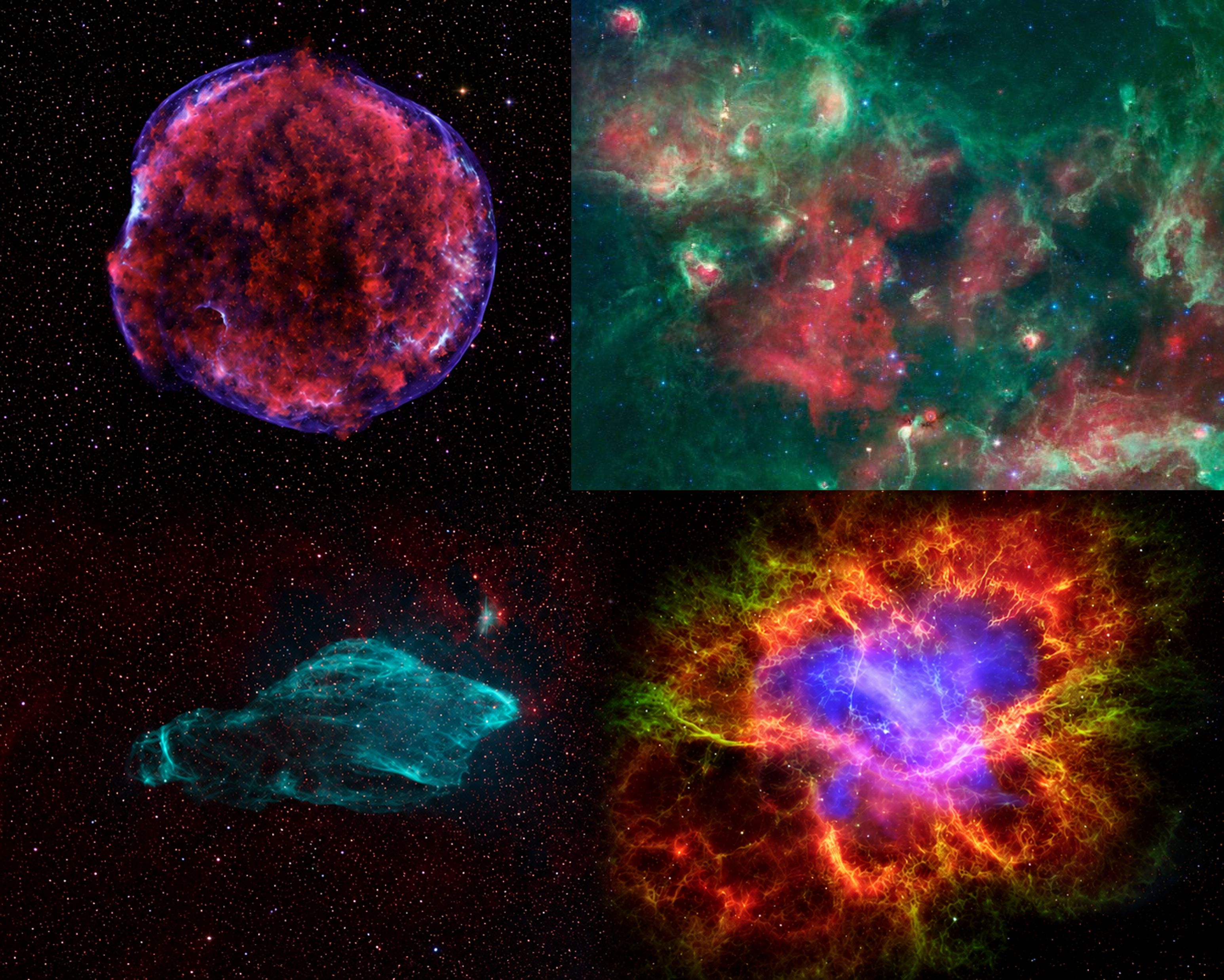}
\caption{
Examples of Galactic PeVatron candidates representing the growing diversity of extreme particle accelerators discovered in recent years. The population includes supernova remnants (e.g.Tycho, Credit: x-ray NASA/CXC/Rutgers/K Eriksen et al; optical: DSS), pulsar wind nebulae (e.g. the Crab, (NASA) Credit: X-ray: NASA/CXC/ASU/J.Hester et al.; Optical: NASA/ESA/ASU/J.Hester \& A.Loll; Infrared: NASA/JPL-Caltech/Univ. Minn./R.Gehrz), star-forming regions (e.g. Cygnus region, Credits: X-Ray: NASA/Spitzer, and microquasar-powered systems (e.g. SS433 in the Manatee Nebula, Credits: NRAO/AUI/NSF, K. Golap, M. Goss; NASA’s Wide Field Survey Explorer (WISE)), highlighting the need for observational diagnostics capable of identifying their particle content. 
}
\label{fig:pevatron_zoo}
\end{figure}
\section{Why TeV-PeV Data Are Not Enough}
\label{sec:TeV}

The discovery of Galactic PeVatrons has transformed the search for the sources responsible for the acceleration of Galactic CRs. While the detection of gamma-rays extending to hundreds of TeV demonstrates the presence of particles accelerated to extreme energies, it does not automatically establish a connection with the Galactic CR population. Since CRs observed near Earth are overwhelmingly composed of hadrons, identifying the sources responsible for the Galactic CR flux requires demonstrating not only that particles are accelerated to PeV energies, but also that these particles are hadronic in nature.

This distinction has become increasingly important as the population of candidate Galactic PeVatrons has diversified (Figure~\ref{fig:pevatron_zoo}). Historically, SNRs were considered the primary candidates for producing Galactic CRs up to the knee through diffusive shock acceleration (DSA) \cite{Hillas05,Blasi13,Bell13,Amato24}. However, recent observations have revealed a much broader population of candidate PeVatrons, including PWNe, YMCs, and MQs \cite{deOnaWilhelmi22,Vieu23,Cardillo23,Kaci25}. These source classes differ substantially in their environments, acceleration mechanisms, and expected particle content, raising the possibility that Galactic PeVatrons may not represent a homogeneous population.

For many years, the detection of photons above $\sim$100 TeV was often regarded as strong evidence for hadronic acceleration. This interpretation was based on the expectation that severe radiative losses would limit the maximum energies achievable by relativistic electrons, making proton acceleration the most plausible explanation for the observed emission \cite{Aharonian04,Hillas05}. Recent observations have challenged this paradigm. A significant fraction of the Galactic PeVatron candidates detected by LHAASO are spatially associated with pulsars and PWNe, environments known to accelerate electrons to extreme energies \cite{Cao24_LHAASO24_cat,deOnaWilhelmi22}. Under realistic physical conditions, inverse Compton scattering can generate gamma-ray spectra extending well into the multi-TeV regime, demonstrating that ultra-high-energy (UHE) photons alone cannot be considered definitive evidence of hadronic acceleration \cite{deOnaWilhelmi22,Amato24}.

The fundamental limitation of TeV-PeV observations is that they primarily constrain the maximum energies reached by accelerated particles, but provide only limited information on their composition. As a consequence, different physical scenarios may produce remarkably similar VHE spectra. This degeneracy is particularly severe in complex Galactic environments where multiple acceleration sites coexist and contribute to the observed emission. Establishing the hadronic nature of candidate PeVatrons is therefore emerging as one of the main challenges in contemporary HE astrophysics.

A representative example is provided by the Cygnus Cocoon, one of the most prominent candidate Galactic PeVatrons. Throughout this work, it will be used as a benchmark for the broader class of Galactic PeVatron candidates. Its combination of bright TeV emission, complex environment, and persistent hadronic/leptonic ambiguity makes it an ideal case study for illustrating the scientific potential of MeV observations. Observations performed by Fermi-LAT, HAWC, LHAASO, and other instruments have established the presence of particles accelerated to at least hundreds of TeV within the Cygnus star-forming region \cite{Ackermann11_Cocoon,Abeysekara21_Cocoon,Cao21}. However, both hadronic and leptonic models can successfully reproduce the currently available TeV data under plausible assumptions regarding particle spectra, radiation fields, and source environments \cite{Aharonian19_Cocoon,Liu24_MeV}. Consequently, despite the exceptional quality of the existing observations, the dominant acceleration mechanism remains uncertain.

This ambiguity is not unique to Cygnus. Similar challenges affect several classes of candidate PeVatrons, including pulsar-powered systems, YMCs, and composite environments hosting multiple particle populations \cite{Cardillo23,Amato24}. As the Galactic PeVatron census continues to expand, the ability to distinguish between hadronic and leptonic accelerators is becoming the primary limitation in interpreting TeV-PeV observations and assessing the contribution of different source classes to the Galactic CR budget.

\begin{figure*}[t]
\centering
\includegraphics[width=0.95\textwidth]{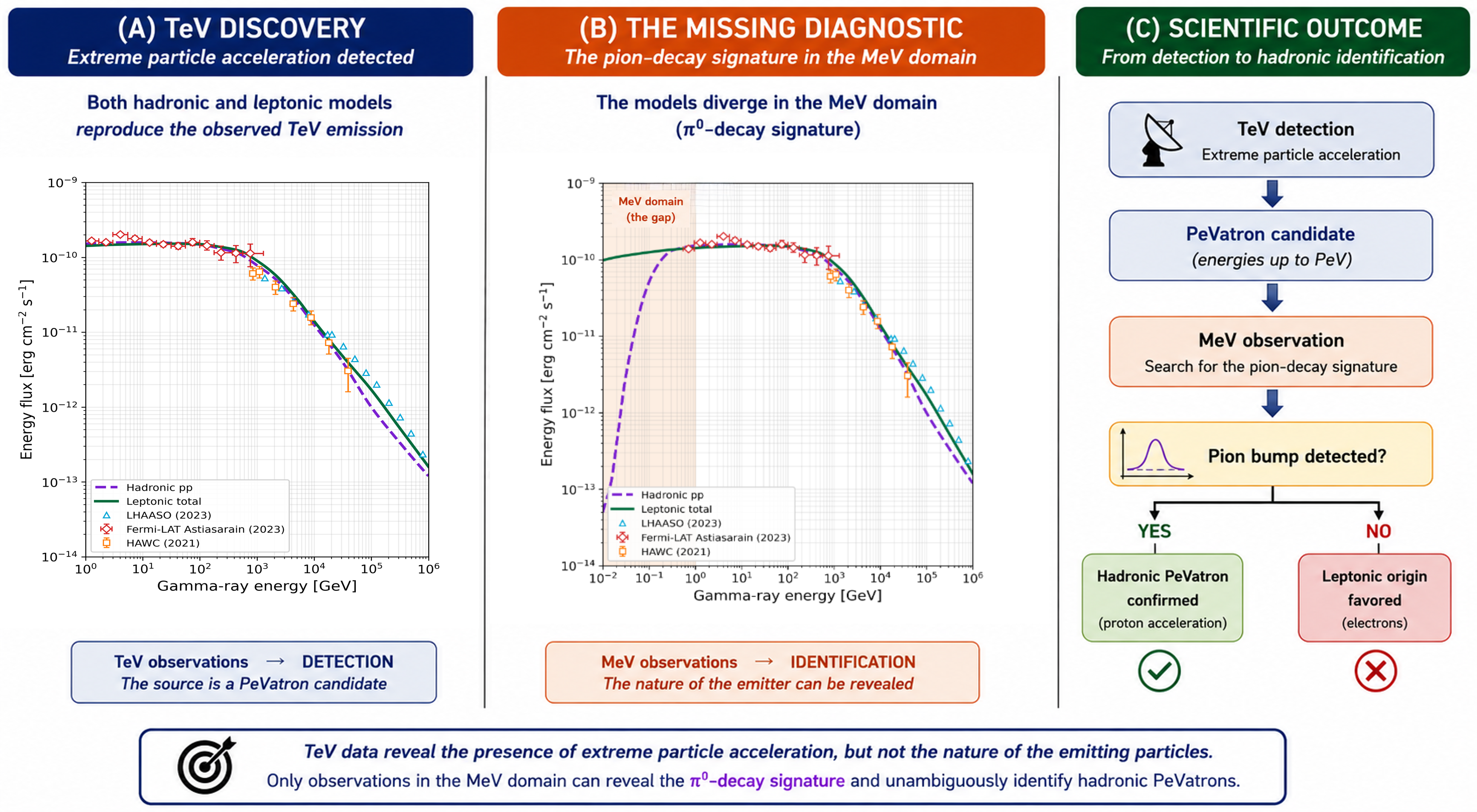}
\caption{
Illustration of the ambiguity affecting the interpretation of Galactic PeVatron candidates. While both hadronic and leptonic scenarios can reproduce the currently available TeV observations of sources such as the Cygnus Cocoon, their predicted behavior differs significantly in the MeV domain. Measurements around the \textit{pion bump} therefore provide a direct means of distinguishing between the two scenarios. 
}
\label{fig:tev_mev}
\end{figure*}

Resolving this ambiguity requires an observable that is directly linked to hadronic interactions rather than to particle energy alone. Such a diagnostic must provide information on the nature of the accelerated particles while remaining robust against the modeling degeneracies that affect broadband TeV observations. As illustrated in Figure~\ref{fig:tev_mev}, measurements in the MeV domain provide exactly this opportunity, since hadronic and leptonic scenarios that appear nearly indistinguishable at TeV energies predict markedly different behaviors around the \textit{pion bump}.

\section{The \textit{pion bump} as a hadronic diagnostic}
\label{sec:pionbump}

When relativistic protons interact with ambient matter, proton-proton collisions produce secondary particles, including neutral pions. These particles decay almost instantaneously into pairs of gamma rays through the process

\[
p + p \rightarrow  p + p + \pi^{0}
\]

\[
\pi^{0} \rightarrow \gamma + \gamma
\]

The resulting gamma-ray spectrum exhibits a characteristic broad feature, commonly known as the \textit{pion bump}, which emerges in the $\sim$10-100 MeV energy range \cite{Dermer86}. As illustrated in Figure~\ref{fig:tev_mev}, hadronic and leptonic scenarios that may appear nearly indistinguishable at TeV-PeV energies predict markedly different behaviors in this energy range. Unlike broadband TeV-PeV emission, whose interpretation often relies on detailed modeling of particle populations and source environments, the \textit{pion bump} is directly linked to hadronic interactions and therefore provides a much more unambiguous tracer of relativistic protons and nuclei.

The diagnostic power of the \textit{pion bump} has already been demonstrated observationally in a limited number of Galactic sources. In particular, AGILE and Fermi-LAT independently reported evidence for the characteristic neutral-pion decay signature in middle-aged SNRs interacting with dense molecular clouds, providing some of the strongest electromagnetic evidence for hadronic particle acceleration in astrophysical environments \cite{Giuliani11,Ackermann13}.

The relevance of this signature extends well beyond the study of individual sources. While leptonic processes such as inverse Compton scattering and Bremsstrahlung can reproduce broad HE gamma-ray spectra under a wide range of physical conditions, they do not naturally generate the characteristic spectral curvature associated with neutral-pion decay \cite{Liu24_MeV}. Consequently, measurements around the \textit{pion bump} provide information on particle composition that is largely inaccessible at higher energies. Even when not sufficient on their own to uniquely determine the origin of the entire gamma-ray spectrum, they can substantially reduce the degeneracies affecting broadband models and provide direct constraints on the hadronic content of the source.

As the population of Galactic PeVatron candidates continues to expand, observations around the \textit{pion bump} offer a unique opportunity to bridge the gap between the discovery of extreme accelerators and the identification of the sources responsible for Galactic CRs. Realizing this scientific potential, however, requires observational capabilities that are currently unavailable in the MeV domain.

\section{The emerging MeV gap}
\label{sec:MeVgap}

\begin{figure*}[t]
\centering
\includegraphics[width=0.95\textwidth]{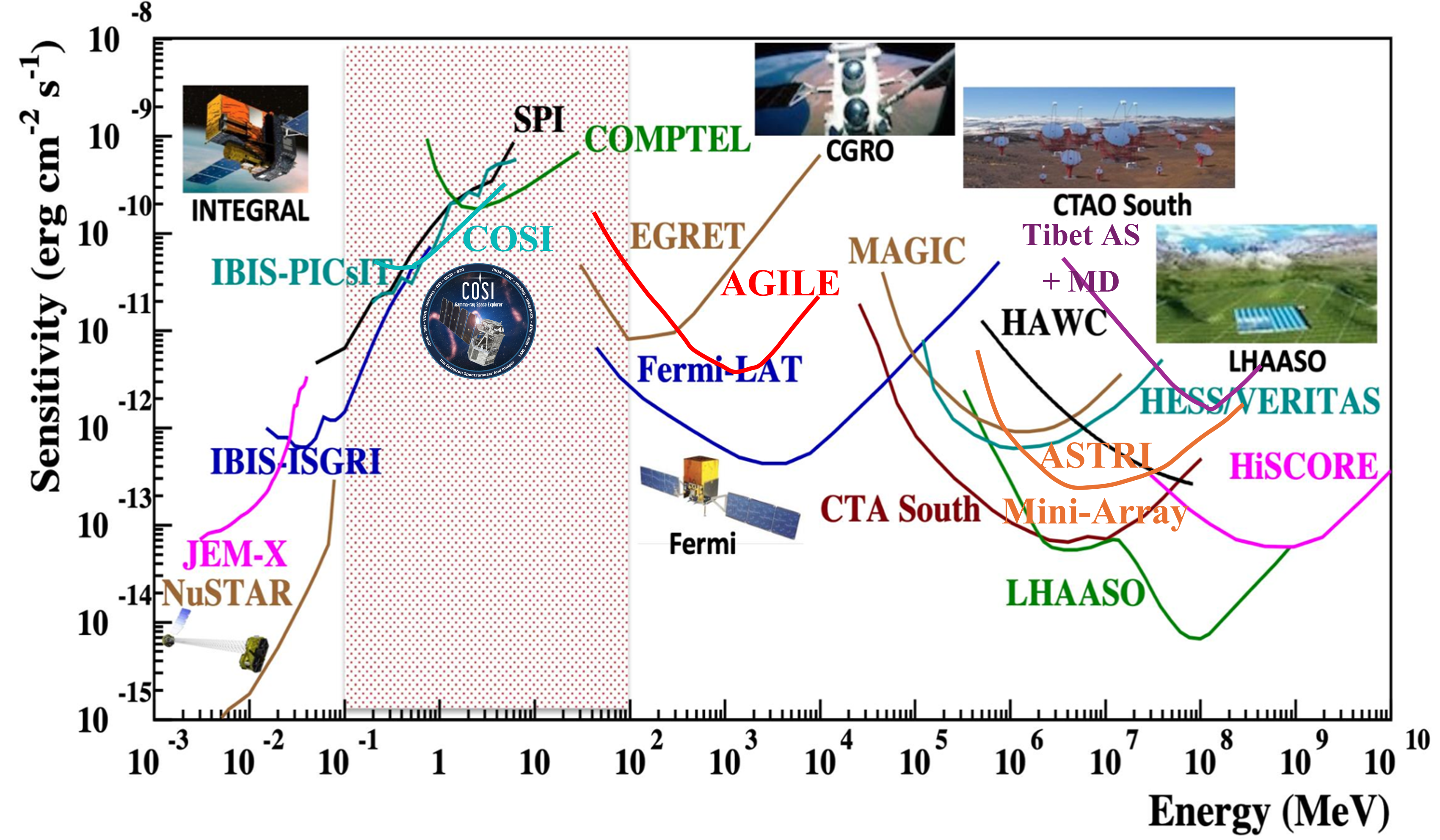}
\caption{
Representative sensitivities of past, current, approved, and proposed MeV gamma-ray observatories. Despite substantial progress in MeV astronomy, a significant observational gap remains in the 10-100 MeV energy range most relevant for systematic \textit{pion-bump} studies of Galactic PeVatron candidates. Image re-adapted from De Angelis et al. 2018\cite{DeAngelis18}}
\label{fig:mev_gap}
\end{figure*}

Despite its scientific importance, the energy range in which it emerges remains one of the least explored regions of the electromagnetic spectrum. This situation reflects a long-standing experimental challenge rather than a lack of scientific interest. Detecting faint Galactic sources in the MeV domain is intrinsically difficult because this energy range combines several unfavorable conditions: the transition between the Compton-scattering and pair-production regimes, large instrumental backgrounds and a bright diffuse Galactic emission that often dominates the observed signal along the Galactic plane \cite{Strong07,Ackermann12,Orlando18,DeAngelis_26}. Together, these factors have historically limited source sensitivity, angular resolution, and source-background discrimination capabilities.

Historically, progress in MeV gamma-ray astronomy has been considerably slower than in the neighboring X-ray, GeV, and TeV domains. While COMPTEL provided the first all-sky view of the MeV sky and demonstrated the scientific richness of this energy range, its sensitivity was insufficient to systematically investigate pion-decay signatures in Galactic particle accelerators \cite{Schonfelder04}. Despite notable successes, including the detection of pion-decay signatures in Galactic SNRs, the observational coverage of the energy range most relevant for pion-decay studies remains limited. This situation is becoming increasingly critical in the era of Galactic PeVatrons. Current and future observatories, including LHAASO, CTAO, and the ASTRI Mini-Array, are expected to continue expanding the population of candidate PeVatrons and to provide increasingly detailed measurements of their VHE emission \cite{Cao24_LHAASO24_cat,CTA19,Scuderi22,Vercellone22}. 

The importance of this observational gap extends beyond the specific science case discussed in this work. The MeV domain addresses a broad range of fundamental astrophysical questions, including CR acceleration and transport, diffuse Galactic emission, compact objects, transients, and multi-messenger studies. Nevertheless, the identification of hadronic particle acceleration in Galactic PeVatrons provides one of the clearest examples of how the lack of sensitive MeV observations is currently limiting scientific progress.

Several next-generation mission concepts have been proposed to address this challenge, including e-ASTRO\-GAM \cite{DeAngelis18,DeAngelis_26}, AMEGO-X \cite{McEnery19_AMEGO}, and ComPair \cite{Moissev15_ComPair,Valverde23_ComPair}, all enabled by major advances in gamma-ray detector technologies. In parallel, the Compton Spectrometer and Imager (COSI), recently selected by NASA as a Small Explorer mission, will provide unprecedented capabilities for studies of nucleosynthesis, Galactic positrons, and compact objects in the $\sim$0.2-5 MeV range \cite{Tomsick19_COSI}. While these efforts represent major advances for MeV astronomy, none is primarily optimized for systematic \textit{pion-bump} studies in Galactic PeVatrons. 

\begin{figure*}[t]
\centering
\includegraphics[width=0.95\textwidth]{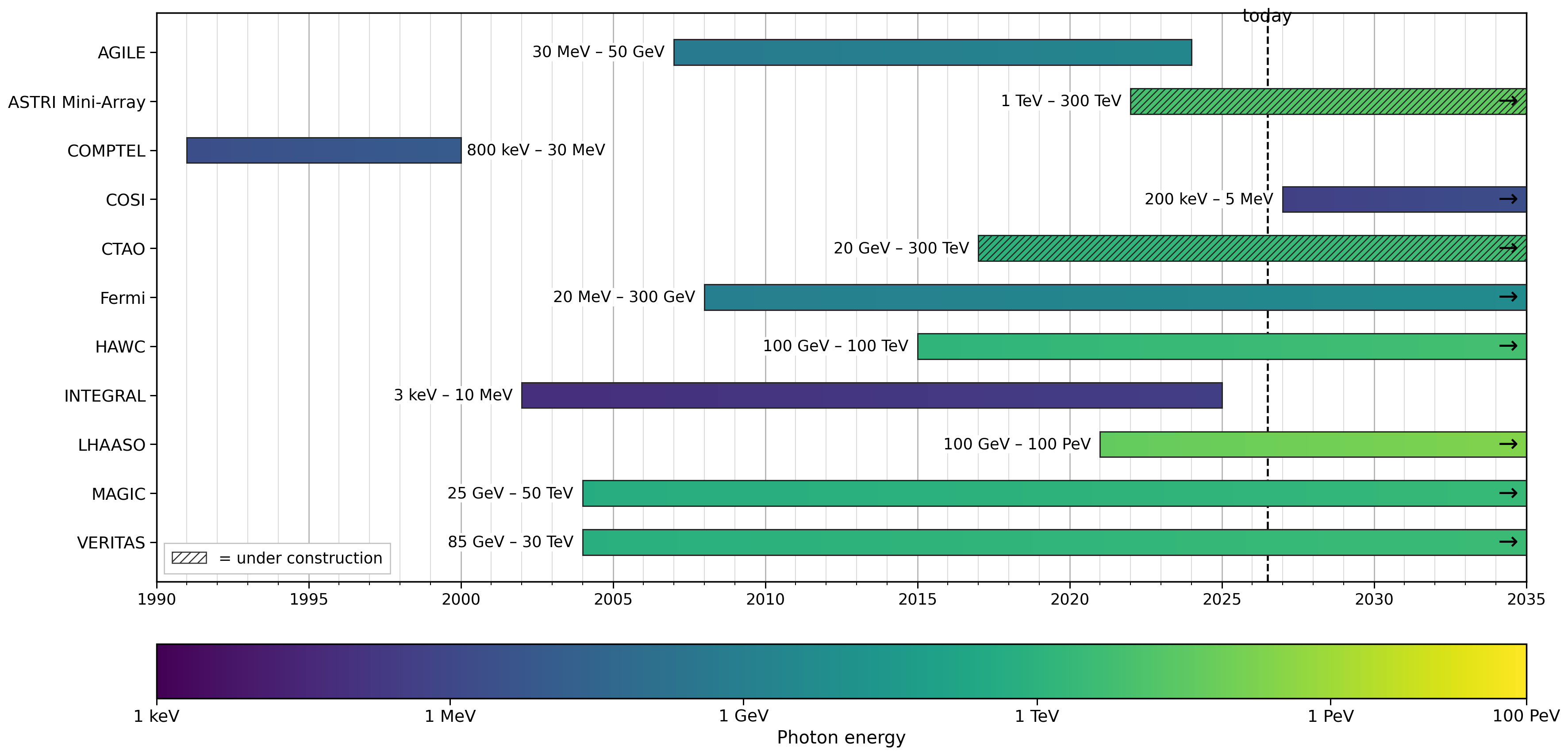}
\caption{
Timeline of major past, present, and future gamma-ray observatories relevant to Galactic PeVatron studies. While the Galactic PeVatron census is expected to expand rapidly through observations by LHAASO, CTAO, and the ASTRI Mini-Array, no approved mission is specifically optimized to investigate the 10-100 MeV energy range where the \textit{pion bump} provides a direct diagnostic of hadronic acceleration.}
\label{fig:why_now}
\end{figure*}

Figure \ref{fig:mev_gap} and Figure \ref{fig:why_now} highlight a growing asymmetry in observational capabilities. The coming decade will be characterized by unprecedented TeV–PeV coverage through LHAASO, CTAO, and the ASTRI Mini-Array, while no comparably mature mission is currently planned for the 10–100 MeV range most relevant to \textit{pion-bump} studies. The challenge is therefore not simply one of sensitivity, but of timing: the capability to discover candidate PeVatrons is advancing faster than our capability to identify their hadronic nature. Addressing this challenge may not require a large general-purpose MeV observatory.

\section{Science Drivers for a Focused MeV Capability}
\label{sec:MeVmission}

The discussion presented in the previous sections suggests that the primary observational challenge is no longer the discovery of additional Galactic PeVatrons, but the identification of their particle content. The objective of a focused MeV capability is therefore not to survey the entire Galactic population, but to obtain decisive measurements for a limited number of benchmark systems capable of addressing the CR origin problem.

To illustrate the origin of the requirements summarized in Table~\ref{tab:req}, we consider the Cygnus Cocoon as a representative Galactic PeVatron candidate. As discussed in Section~\ref{sec:TeV}, the Cocoon provides an ideal benchmark because, despite the exceptional quality of the currently available GeV-TeV observations, both hadronic and leptonic scenarios remain viable explanations for the observed emission. Recent studies have shown that the MeV domain hosts the transition between pion-decay and Bremsstrahlung emission, with the two components becoming increasingly distinguishable below $\sim100$ MeV and the $\sim10$-50 MeV range providing particularly strong diagnostic power for hadronic/leptonic discrimination \cite{Liu24_MeV,Peron26}.

The Cygnus Cocoon should be regarded as an optimistic benchmark rather than as a typical target. It is among the brightest and best-studied Galactic PeVatron candidates, with extensive multi-wavelength and multi-TeV coverage. Most of the currently known or expected PeVatron candidates are likely to be fainter in the MeV domain, more confused, or less well constrained environmentally. Therefore, requirements derived from the Cocoon do not represent an over-optimization for a single favorable source, but rather a lower limit on the capabilities needed to perform \textit{pion-bump} studies on the broader Galactic PeVatron population.

The first key requirement is sensitivity across the energy range where the \textit{pion bump} emerges. Reliable measurements from approximately 10 MeV to a few hundred MeV are required to detect the characteristic spectral curvature and to distinguish hadronic emission from competing leptonic scenarios. Because candidate Galactic PeVatrons are expected to be faint in this energy range, the most effective observing strategy is not an all-sky survey but deep pointed observations of a limited number of carefully selected targets. This strategy follows the logic that led to the identification of the \textit{pion bump} in bright middle-aged SNRs such as W44. The objective is not to classify every Galactic PeVatron candidate, but to obtain decisive measurements for a limited number of benchmark systems where the scientific return is highest and the implications for the CR origin problem are most significant. In this sense, the detection or exclusion of a \textit{pion-bump} signature in the Cygnus Cocoon would represent a critical proof of concept: success on this benchmark would demonstrate that similar measurements may become feasible for fainter or less favorable PeVatron candidates, while failure would immediately quantify how demanding systematic \textit{pion-bump} studies will be.

Achieving the required sensitivity, however, is meaningful only if the instrument can also separate source emission from the surrounding Galactic environment. Many candidate PeVatrons, including the Cygnus Cocoon, are extended systems embedded in regions characterized by strong diffuse emission and multiple potential particle accelerators. Sensitivity and angular resolution therefore become tightly coupled requirements. Adequate angular resolution is needed not only to associate the MeV emission with the correct TeV counterpart, but also to limit contamination from nearby sources and structured diffuse backgrounds, particularly along the Galactic plane \cite{Strong07,Ackermann12,Orlando18}.

A further requirement is strong background control. In the 10-300 MeV range, instrumental background, Earth albedo, and diffuse Galactic emission can significantly affect the detectability of weak spectral features. The relevant performance metric is therefore not only continuum sensitivity, but the capability to recover spectral curvature under realistic background conditions. This naturally favors a science-driven optimization of event reconstruction, background rejection, observing strategy, and target selection.

These requirements, summarized in Table \ref{tab:req}, do not necessarily demand a general-purpose MeV observatory. Rather, they motivate a compact and dedicated mission concept optimized for deep observations of a limited sample of benchmark PeVatron candidates. Recent advances in silicon-based trackers, scintillating fibers, silicon photomultipliers, and low-power readout electronics have expanded the range of feasible detector architectures for MeV gamma-ray astronomy \cite{DeAngelis18,McEnery19_AMEGO,Valverde23_ComPair}. Multiple implementation pathways may be envisaged, spanning different mission classes and levels of complexity; the present discussion is intentionally science-driven and does not assume a specific mission architecture. Even a limited number of robust detections, or physically constraining non-detections, would provide crucial information on the hadronic content of Galactic accelerators and guide the science case of future large-scale MeV observatories. The goal of Table \ref{tab:req} is not to define mission requirements with engineering precision, but to identify the observational parameter space within which \textit{pion-bump} studies of Galactic PeVatrons become scientifically viable.

\begin{table}[ht]
\centering
\caption{Science-driven performance requirements for a focused \textit{pion-bump} mission. The values should be regarded as indicative targets derived from benchmark science cases and from the capabilities of past, current, and proposed MeV observatories, rather than as the result of a formal mission optimization study.}
\vspace{6pt}
\label{tab:req}
\begin{tabular}{p{3.2cm} p{2.8cm} p{7.0cm}}
\hline
\textbf{Parameter} & \textbf{Target value} & \textbf{Scientific motivation} \\
\hline
Energy range &
10--300 MeV &
Coverage of the neutral-pion decay feature and connection with the GeV domain for hadronic/leptonic discrimination. \\

Continuum sensitivity &
$\sim10^{-11}$ erg cm$^{-2}$ s$^{-1}$ &
Detailed pion-decay studies of benchmark systems such as the Cygnus Cocoon and meaningful constraints for the broader population of fainter Galactic PeVatron candidates. \\

Angular resolution &
$\lesssim 1^\circ$--$2^\circ$ &
Separation of overlapping emission components in complex Galactic environments and association of the MeV emission with the correct TeV counterpart.\\

Background rejection &
Science-driven optimization &
Control of instrumental background, Earth albedo, and diffuse Galactic emission, which dominate the uncertainty budget in the 10--300 MeV domain. \\

Energy resolution &
$\lesssim 20\%$ &
Identification of the characteristic shape of the pion-decay feature and discrimination from competing leptonic models. \\

Observing strategy &
Deep pointed observations &
Long integrations on a small sample of carefully selected PeVatron candidates rather than all-sky population surveys. \\

Mission profile &
Compact / fast-track compatible &
Rapid response to the emerging PeVatron landscape while complementing future large-scale MeV observatories. \\
\hline
\end{tabular}
\end{table}

While the detection of pion-decay emission would provide compelling evidence for hadronic interactions, not all Galactic PeVatron candidates are necessarily expected to exhibit a detectable \textit{pion bump}. Low ambient densities, particle escape, source evolution, and line-of-sight confusion may reduce the observable hadronic signature even in the presence of efficient proton acceleration. Consequently, non-detections can be scientifically valuable, as they provide constraints on the environments and acceleration conditions of candidate PeVatrons. A dedicated MeV capability would therefore serve not only as a discovery instrument, but also as a critical tool for testing and refining theoretical models of Galactic particle acceleration.

\section{Conclusions}
\label{sec:conclusions}

The rapidly growing population of Galactic PeVatron candidates is transforming the study of extreme particle accelerators from the investigation of a handful of exceptional objects into a true population science. Observatories such as LHAASO have already revealed a rich and diverse Galactic PeVatron landscape, while CTAO and the ASTRI Mini-Array are expected to further expand the census and provide increasingly detailed measurements of particle acceleration up to PeV energies.

As this observational revolution unfolds, the central challenge is shifting from source discovery to source identification. Establishing whether candidate PeVatrons are accelerating hadrons or leptons is now one of the key limitations in understanding the origin of Galactic CRs. In many cases, TeV-PeV observations alone cannot provide a definitive answer, as fundamentally different physical scenarios can produce remarkably similar VHE gamma-ray signatures.

Measurements around the \textit{pion bump} offer a unique opportunity to overcome this limitation. As the most direct electromagnetic tracer of hadronic interactions, the neutral-pion decay feature provides the missing observational link between the detection of extreme particle accelerators and the identification of genuine Galactic CR sources. At the same time, the energy range where this diagnostic emerges remains one of the least explored regions of the electromagnetic spectrum.

This situation creates a growing mismatch between scientific opportunity and observational capability. While the Galactic PeVatron census is expected to increase dramatically over the coming decade, no approved mission is specifically optimized to investigate the 10-100 MeV energy range where hadronic and leptonic scenarios diverge most strongly. Without sensitive observations in this domain, the scientific return of the ongoing PeVatron revolution may remain fundamentally incomplete.

Recent advances in detector technologies make it increasingly realistic to consider compact and focused mission concepts dedicated to deep observations of benchmark Galactic PeVatron candidates. Such a capability would complement future large-scale MeV observatories while providing a timely response to one of the most pressing observational challenges in high-energy astrophysics.

We are discovering PeVatrons faster than our ability to identify them. Resolving the origin of Galactic CRs may ultimately require bringing the \textit{pion bump} back to the center of gamma-ray astronomy.

\bibliography{report} 
\bibliographystyle{spiebib} 

\end{document}